\documentclass[preprint]{JASA}

\makeatletter
\let\linenumbers\relax

\makeatother

\usepackage{ccaption}
\usepackage{multirow}
\begin{document}

\title[Qian et al./SPMnet for Automotive sound field reproduction]{Automotive sound field reproduction using deep optimization with spatial domain constraint}
\author{Yufan Qian}
\author{Xihong Wu}
\author{Tianshu Qu}

\affiliation{State Key Laboratory of General Artificial Intelligence, School of Intelligence Science and Technology, Peking University, Beijing, 100871, China}

 

\preprint{Yufan Qian, Tianshu Qu, Xihong Wu, JASA}	

\date{\today}

\begin{abstract}
Sound field reproduction with undistorted sound quality and precise spatial localization is desirable for automotive audio systems. However, the complexity of automotive cabin acoustic environment often necessitates a trade-off between sound quality and spatial accuracy. To overcome this limitation, we propose Spatial Power Map Net (SPMnet), a learning-based sound field reproduction method that improves both sound quality and spatial localization in complex environments. We introduce a spatial power map (SPM) constraint, which characterizes the angular energy distribution of the reproduced field using beamforming. This constraint guides energy toward the intended direction to enhance spatial localization, and is integrated into a multi-channel equalization framework to also improve sound quality under reverberant conditions.
To address the resulting non-convexity, deep optimization that use neural networks to solve optimization problems is employed for filter design. Both \textit{in situ} objective and subjective evaluations confirm that our method enhances sound quality and improves spatial localization within the automotive cabin. Furthermore, we analyze the influence of different audio materials and the arrival angles of the virtual sound source in the reproduced sound field, investigating the potential underlying factors affecting these results.


\end{abstract}
  

\maketitle


\section{\label{sec:1} Introduction}
The automotive cabin is increasingly used for enjoying high-quality audio. With the growing number of loudspeakers in the cabin, the rising popularity of object-based audio contents such as Dolby Atmos, and the advancement of quieter electric engines, the experience of immersive spatial audio within the cabin has become progressively more feasible. To achieve this, sound field reproduction techniques are typically required.

Over the past few decades, numerous methods for sound field reproduction using loudspeaker arrays have been proposed, including Wave Field Synthesis (WFS)\cite{berkhout1993wfs}, Higher Order Ambisonics (HOA) \cite{ward2001reproduction}, and Vector-based Amplitude Panning (VBAP) \cite{pulkki1997vbap}. These approaches typically assume that the loudspeaker array consists of loudspeakers with identical frequency responses, arranged according to specific geometric rules (e.g., uniformly distributed on a sphere) in an anechoic environment \cite{Poletti2005hoa}. However, in a practical automotive cabin, complex sound reflections and absorption due to windows and interior materials, the variability in loudspeaker frequency responses, and the irregular arrangement of the loudspeaker array violate these assumptions \cite{kleiner1998objective, house1989aspects, cheer2019sound}, resulting in degraded sound field reproduction.

Acoustic channel equalization, which utilizes IRs between loudspeakers and control points to compensate for the effects of the enclosure, is a major approach to addressing these challenges \cite{Cecchi2017RRE}. Traditionally, methods based on spatial sampling of the reproduction zone and direct least-squares equalization of the acoustic channel have been employed \cite{Kirkeby1998, Kirkeby_Nelson_1999}. However, these approaches often suffer from poor spatial robustness \cite{Fielder_2001}, leading to degraded reproduction performance, including compromised sound quality and spatial localization between control points. For robust reproduction, several methods aligns the IRs of multiple control points, followed by clustering or prototype extraction to obtain a spatially smoothed IR of the environment\cite{omiciuolo2008multiple, Cecchi2009}. However, these approaches blur the relative phase characteristics among control points, resulting in imprecise spatial localization\cite{walker1998equalization}. Additionally, if proper time-domain constraints are not incorporated during filter design, pre-ringing artifacts can further degrade sound quality\cite{Elliott_Nelson_1989}. 

WFS and HOA-based equalization techniques have been proposed\cite{Corteel_2006, Gauthier_Berry_2006,  Betlehem_Abhayapala_2005, Lecomte_Gauthier_Langrenne_Berry_Garcia_2018, ge2021partially}, leveraging the Fourier-Bessel representation of the sound field for reproduction over an extended area within the enclosure\cite{rafaely2015fundamentals, zotter2019ambisonics}. While these methods preserve spatial localization and achieve a flat frequency response, the equalized IRs inherently suffer from uncontrolled pre-ringing artifacts due to the non-minimum phase nature of acoustic channels\cite{Neely_Allen_1979}, ultimately degrading sound quality\cite{Dolan_Small_1984}. Although minimum phase equalization filters could be used as a complement, their degraded performance was not preferred\cite{maamar2006partial}. Furthermore, these methods still rely on a relatively well-structured loudspeaker array with consistent frequency responses, which contradicts the practical conditions in an automotive environment. 

To control pre-ringing artifacts during reproduction, polynomial-based methods were utilized in \cite{Brannmark_Ahlen_2009, Brannmark2013}, attempting to extract common zeros from IRs across multiple control points to design mixed-phase equalization filters without introducing pre-ringing artifacts. Envelope control of equalized IRs has also been proposed in \cite{qian2024automotive}, designing filters that introduce only inaudible pre-ringing artifacts. Although these methods managed to equalize the non-minimum IR in the automotive, reproducing the sound field with unimpaired sound quality, the spatial localization is still limited because they do not explicitly impose constraints on spatial localization but instead rely on the implicit constraints derived from the amplitude and phase characteristics of IRs of multiple control points.

In this paper, we propose an automotive sound field reproduction method named SPMnet, which is capable of render virtual sound images with impaired sound quality and accurate spatial localization. The contributions of this work are three-fold. Firstly, we consider both sound quality and spatial localization as performance evaluation metrics for sound field reconstruction in the automotive cabin. Secondly, a spatial power map constraint based on beamforming is introduced and incorporated to explicitly control the spatial localization of the reproduced sound field. Lastly, deep optimization techniques were employed to address the potential non-convexity of the newly proposed constraint.

The remainder of this paper is organized as follows. Section \ref{sec:2 method} describes our proposed method, including the problem formulation, conventional approach, and our proposed SPMnet method. Section \ref{sec3: objEval} presents the objective experiment results. Section \ref{sec4: subEval} details the experimental design and outcomes of two subjective listening tests. Section \ref{sec5: discuss} discusses the consistencies and discrepancies between the objective and subjective results and explores their potential causes. Finally, Section \ref{sec6: conclusion} concludes the paper and outlines future work.

\section{\label{sec:2 method} Method}
\subsection{\label{sec2sub1: overall} Problem Formulation}
\begin{figure}
    \centering
    \includegraphics[width=0.7\linewidth]{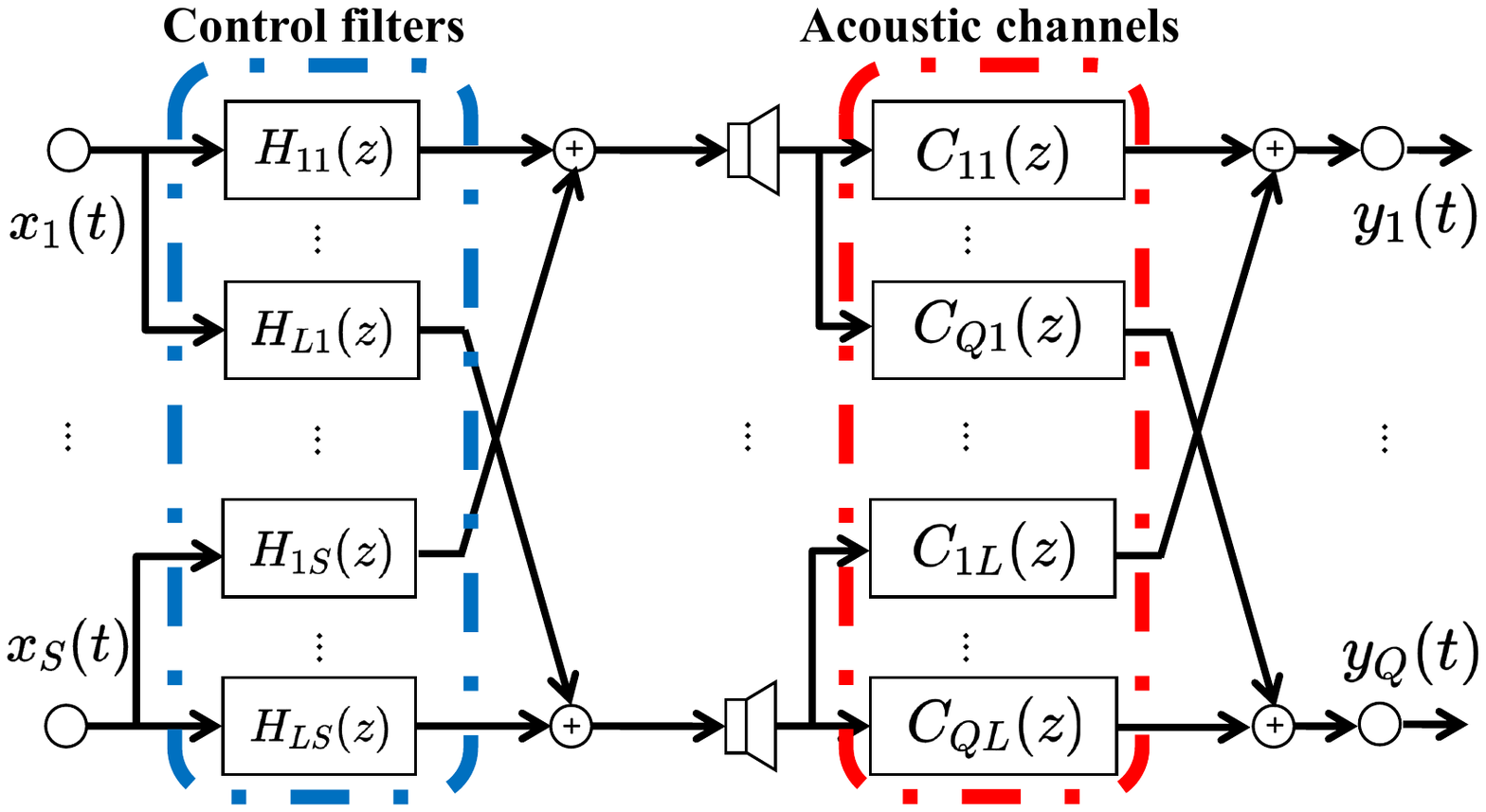}
    \caption{Sound field reproduction system structure consisting of control filters and acoustic channels}
    \label{fig:overall system}
\end{figure}
The primary objective of this study is to achieve sound field reproduction around the driver's head position that simultaneously ensures high sound quality and accurate spatial localization, using only the original loudspeaker array integrated into the vehicle.  To this end, we define the sound field reproduction system for $S$ virtual sources to be reproduced, $L$ loudspeaker channels and $Q$ microphones around the head position as a linear time-invariant (LTI) system as depicted in Fig.~\ref{fig:overall system}. Assuming FIR filters, acoustic channel $c_{ql}(t)$ of length $L_c$ and control filters $h_{ls}(t)$ of length $L_h$, the global response is given by 
\begin{equation}\label{eq:global impulse response convolution}
    g_{qs}\left( t \right) =\sum_{l=1}^L{c_{ql}\left( t \right) *h_{ls}\left( t \right)}, s= 1,\cdots,S, q=1,\cdots,Q, 
\end{equation}
with $g_{qs}(t)$ indicating the global impulse response from source $s$ to microphone $q$ of length $L_g=L_c+L_h-1$. As control filters can be designed independently for each of the $S$ sources, single source assumption is made for simplicity. The impulse response $d_{qs}(t)$ of length $L_g$ which represents the acoustic transfer from source $s$ to microphone $q$ in an anechoic room is introduced as an ideal transmission target. The ultimate purpose of the reproduction system is to force the actual response $g_{qs}(t)$ to mitigate the target response as 
\begin{equation}\label{eq:ultimate purpose }
   g_{qs}(t) = d_{qs}(t).
\end{equation}
By vectorizing the aforementioned IRs (e.g., $\mathbf{h}_{ls} = [h_{ls}(0), \cdots, h_{ls}(L_h-1)]^T$), and denoting the $L_g \times L_c$ convolution matrix constructed from the acoustic channel $c_{ql}(t)$ as $\mathbf{C}_{ql}$, we can incorporate all $Q$ microphones by concatenating the individual responses as
\begin{align*}\label{D h C}
    \mathbf{D}_s=\left[ \begin{array}{c}
	\mathbf{d}_{1s}\\
	\vdots\\
	\mathbf{d}_{Qs}\\
    \end{array} \right] ,
    \mathbf{h}_s=\left[ \begin{array}{c}
	\mathbf{h}_{1s}\\
	\vdots\\
	\mathbf{h}_{Ls}\\
    \end{array} \right] ,
    \mathbf{C}=\left[ \begin{matrix}
	\mathbf{C}_{11}&		\cdots&		\mathbf{C}_{1L}\\
	\vdots&		\ddots&		\vdots\\
	\mathbf{C}_{Q1}&		\cdots&		\mathbf{C}_{QL}\\
    \end{matrix} \right]
\end{align*}
This allows us to formulate the design of the control filters as the following optimization problem:
\begin{equation}\label{eq:matrix global impulse response convolution}
    \mathbf{h}_{s}^{opt}=\underset{\mathbf{h}_{s}}{arg\min}\Delta(\mathbf{Ch}_s, \mathbf{d}_s),
\end{equation} 
where $\Delta(\cdot , \cdot )$ is the objective function defined to quantify the deviation between the reproduced and the desired sound field. The choice of objective function and optimization algorithm plays a critical role in the filter design. In particular, while the objective function determines which aspects of the reproduced sound field (e.g., frequency flatness, spatial fidelity) are prioritized, the optimization algorithm affects whether the solution can be efficiently and reliably obtained. These two factors jointly shape the practical effectiveness of sound field reproduction.

\subsection{Conventional Approach}\label{sec2sub2: bSPM}
To analyze the limitations of existing reprouction methods, we consider two representative conventional approaches—Frequency Deconvolution (FD) \cite{Kirkeby1998} and a convex formulation method (CVX)\cite{qian2024automotive}—which are also used as baselines later in experiments.
The FD method defines its objective function in the frequency domain. This function integrates a least mean square (LMS) error term with a high-frequency regularization term, and the resulting optimization problem is solved analytically.  While effective in flattening the frequency response, it introduces pre- and post-ringing artifacts in the time domain and suffers from limited spatial robustness. The CVX method, by contrast, formulates the filter design mostly follow the concept of temporal shaping and solves the resulting convex optimization problem using an adaptive proximal gradient algorithm. The convex objective function acts as 
\begin{equation}\label{eq:overall objective function}
\begin{aligned}
   \mathbf{h}_{s}^{opt}=\underset{\mathbf{h}_{s}}{arg\min}&\frac{\lambda _1}{2}\left\| \mathbf{W}_{s}^{d}\left( \mathbf{Ch}_s-\mathbf{d}_s \right) \right\| _{2}^{2}+\frac{\lambda _2}{2}\left\| \mathbf{W}_{s}^{u}\mathbf{Ch}_s \right\| _{\infty}\\
    +&\frac{\lambda_3}{2}\left\|\mathbf{V}_s\odot(\widetilde{\mathbf{F}}\mathbf{Ch}_s) \right\|_{\infty}+\frac{\lambda_4}{2}\left\|\mathbf{\xi}\odot(\widetilde{\mathbf{F}}\mathbf{h}_s) \right\| _{\infty},
\end{aligned}
\end{equation}
where the hyperparameters $\lambda_1$ to $\lambda_4$ balance the four terms of the objective. $\mathbf{W}{s}^{d}$ and $\mathbf{W}{s}^{u}$ are window functions that extract and weigh the desired (direct path) and unwanted (pre- and post-ringing artifacts) parts of the global response, respectively. $\widetilde{\mathbf{F}}$ denotes the discrete Fourier transform (DFT) matrix with appropriate dimensions, and $\odot$ represents the element-wise (Hadamard) product. $\mathbf{V}s$ defines the target spectral shape (a flat response with constant weights in our case), and $\boldsymbol{\xi}$ denotes a working frequency-band mask. Notation $|\cdot|_2$ and $|\cdot|_{\infty}$ indicate the $\ell_2$ and $\ell_{\infty}$ norms, respectively. 
In general, the terms weighted by $\lambda_1$ and $\lambda_2$ represent temporal constraints, while those weighted by $\lambda_3$ and $\lambda_4$ correspond to frequency-domain constraints. All four terms were designed with the goal of improving sound quality. We refer to each term by the subscript index of its respective $\lambda$:
\begin{itemize}
    \item Term $\lambda_1$: Minimizes the peak error between the actual global response and the target response. In multi-microphone setups, it also implicitly aligns temporal structure and relative amplitude, benefiting spatial localization.
    \item Term $\lambda_2$: Constrains ringing artifacts within a predefined temporal envelope in $\mathbf{W}{s}^{u}$, exploiting auditory temporal masking to suppress perceptual distortions.
    \item Term $\lambda_3$: Enforces frequency-domain flatness in the reproduced global response.
    \item Term $\lambda_4$: Restricts the effective bandwidth of control filters to avoid driving loudspeakers beyond their intended range.
\end{itemize}
Although the $\lambda_1$ term implicitly contributes to spatial localization by aligning the temporal peaks across microphones, its effectiveness is limited (see comparison in Sec. \ref{sec3sub3: dirAnalyze}). This is because such methods are fundamentally designed for acoustic channel equalization, an objective that prioritizes sound quality over spatial fidelity. Consequently, they lack explicit spatial constraints, leading to poor localization that is inconsistent with our primary goal.

\subsection{Proposed Approach}\label{sec2sub3: SPMnet}
\subsubsection{Spatial domain constraint}\label{sec2sub3sub1: SPMnet}
Instead of relying on existing temporal or frequency-domain constraints, which only implicitly affect spatial localization, we propose the spatial power map (SPM): a constraint applied directly in the spatial domain to explicitly improve localization. This constraint aims to guide the angular distribution of the acoustical energy of the reproduced sound field toward the desired direction. While high-resolution spatial power map estimation methods such as MVDR and MUSIC exist\cite{souden2010study, schmidt1986multiple}, we estimate the SPM using delay-and-sum beamforming (DSB), which offers a favorable balance between spatial resolution and robustness \cite{krishnaveni2013beamforming}. The beamforming process is calculated as:
\begin{equation}\label{eq:spatial power map}
  \Gamma_{bs} = \sum_{f=0}^{F-1} \left[\sum_{q=1}^{Q} \tilde{g}_{qs}(f) \tilde{\omega}_{qb}(f) \right]^2, \quad b = 1, \cdots, B
\end{equation}
where $B$ is the number of steering directions, and $F$ is the number of frequency bins. The term $\tilde{g}_{qs}(f)$ denotes the discrete Fourier transform (DFT) of the global response $g_{qs}(t)$, while $\tilde{\omega}_{qb}(f)$ denotes the beamforming weight at microphone $q$ for steering angle $b$, determined by the microphone position and the desired steering direction~\cite{krishnaveni2013beamforming}. 
Through concatenating of all $B$ scanning angles and vectorization of $\tilde{g}_{qs}(f)$ and $\tilde{\omega}_{qb}(f)$, similar to the process above, the estimated SPM goes as 
\begin{equation}\label{eq:matrix spatial power map}
  \mathbf{\Gamma}_{s} = \mathbf{\Omega} \tilde{\mathbf{G}}_s,
\end{equation}
where
\begin{align*}\label{Gamma Omega G}
 \mathbf{\Gamma}_s=\left[ \begin{array}{c}
	\Gamma_{1s}\\
	\vdots\\
	\Gamma_{Bs}\\
    \end{array} \right] ,
 \tilde{\mathbf{G}}_s=\left[ \begin{array}{c} 
 \tilde{\mathbf{g}}_{1s}\\
	\vdots\\
	\tilde{\mathbf{g}}_{Qs}\\
    \end{array} \right] ,
 \mathbf{\Omega}=\left[ \begin{matrix}
 \mathbf{\tilde{\omega}}_{11}&\cdots&\mathbf{\tilde{\omega}}_{Qb}\\
 \vdots&		\ddots&		\vdots\\
 \mathbf{\tilde{\omega}}_{1B}&\cdots&\mathbf{\tilde{\omega}}_{QB}
    \end{matrix} \right].
\end{align*}
Together with the target SPM $\mathbf{\tilde{\Gamma}}_s$ estimated using $\mathbf{D}_s$, the SPM constraint is defined as
\begin{equation}\label{eq:spatial power map constraint}
    \Delta_{SPM} = \left \| \mathbf{\Gamma}_{s} - \mathbf{\tilde{\Gamma}}_s \right \| _2^2.
\end{equation}
It’s worth noting that our choice of DSB for estimating the SPM is based on two main factors. First, DSB, as a signal-independent beamforming method, is computationally efficient and stable. Unlike signal-dependent methods such as MVDR or MUSIC, which require matrix inversion and may introduce numerical risks that could lead to divergence, DSB only requires matrix multiplication, ensuring a stable and fast calculation. Second, DSB offers a balanced resolution with strong robustness, making it less prone to overfitting during optimization. Additionally, considering the limited spatial localization capability of the human auditory system \cite{letowski2012auditory}, the perceptual improvements gained from an excessively high-resolution SPM estimator may be minimal.
Although incorporating the introduced SPM constraint into Equ.~(\ref{eq:overall objective function}) results in a new objective function that directly considers spatial localization, finding an appropriate optimization algorithm to solve it remains a challenge, as the objective function has been significantly altered in terms of its convexity.

\subsubsection{Deep optimization}\label{sec2sub3sub2: deep optimization}
\begin{figure}
    \centering
    \includegraphics[width=1\linewidth]{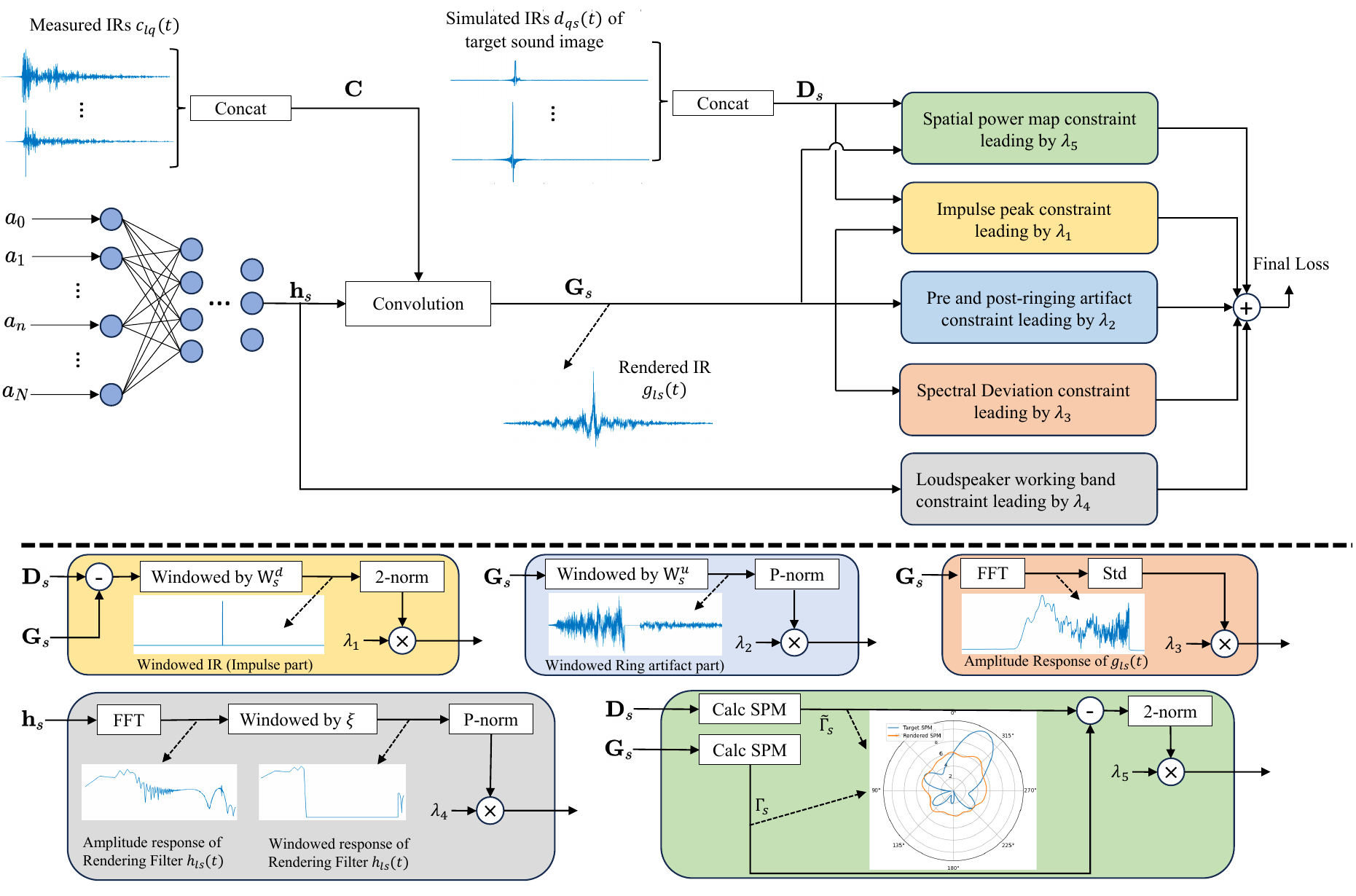}
    \caption{SPMnet overview: the network outputs control filter coefficients $\mathbf{h_s}$ utilized for calculating objective function. The calculation process and intermediate results of all five different objective functions are separately detailed at the bottom of the figure. After obtaining the final loss function, back propagation is used to train the network for the optimal network parameters. The inputs $a_0$ to $a_N$ are random numbers but stay unchanged during training.}
    \label{fig:SPM-Net}
\end{figure}

While the impact of incorporating the SPM constraint remains to be fully demonstrated, the inherent non-convexity of this constraint renders conventional convex optimization-based filter design methods unsuitable. Specifically, the non-convexity stems from the squared summation involved in computing the beamforming output in Equ.~(\ref{eq:spatial power map}), as this is a nonlinear operation. We chose to explore Deep Optimization, a deep learning-based approach for non-convex optimization, given its proven effectiveness in simplifying such problems \cite{lopez2018easing}. Inspired by \cite{Pepe2020FIR, Pepe2022IIR}, which utilized deep learning for the design of IIR and FIR filters, we extended this approach to solve our proposed objective function. With the defined objective function acting as the loss function, the neural network is trained to perform the non-convex optimization task. We name after the proposed method as Spatial Power Map net (SPMnet) and its overall structure is illustrated in Fig.\ref{fig:SPM-Net}. 
The new objective function is calculated as
\begin{equation}\label{eq:new overall objective function}
\begin{aligned}
   \mathbf{h}_{s}^{opt}=\underset{\mathbf{h}_{s}}{arg\min}&\frac{\lambda _1}{2}\left\| \mathbf{W}_{s}^{d}\left( \mathbf{Ch}_s-\mathbf{d}_s \right) \right\| _{2}^{2}+\frac{\lambda _2}{2}\left\| \mathbf{W}_{s}^{u}\mathbf{Ch}_s \right\| _{p}\\
    +&\frac{\lambda_3}{2}std(20log_{10}(\widetilde{\mathbf{F}}\mathbf{Ch}_s))+\frac{\lambda_4}{2}\left\|\mathbf{\xi}\odot(\widetilde{\mathbf{F}}\mathbf{h}_s) \right\| _{p}\\
    +&\frac{\lambda_5}{2}\left \| \mathbf{\Gamma}_{s} - \mathbf{\tilde{\Gamma}}_s \right \| _2^2.
\end{aligned}
\end{equation}
Beyond incorporating the new SPM constraint, weighted by $\lambda_{5}$, we introduce two further modifications. First, all infinity norms ($\ell_\infty$-norm) are approximated by a differentiable large-$\ell_p$ norm. This approximation is essential for ensuring the differentiability required by the backpropagation algorithm during neural network training. Second, the frequency-domain flatness constraint (governed by $\lambda_3$) is substituted with a spectral deviation constraint. This change is motivated by the fact that infinity-norm-based constraints exclusively penalize peaks in the magnitude response, while ignoring notches. Our revised two-sided constraint addresses both simultaneously. Notably, this formulation was avoided in prior research, which was restricted to convex objective functions \cite{ Jungmann_Mei_Goetze_Mertins_2011}.

\section{\label{sec3: objEval} Objective Experiment}
\subsection{\label{sec3sub1: cabin} Experimental Setup}
\begin{figure}
    \centering
    \includegraphics[width=0.7\linewidth]{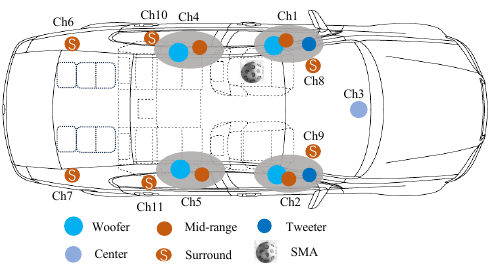}
    \caption{Loudspeaker arrangement inside the automotive cabin. Gray ellipses indicate loudspeaker units (woofer, mid-range, tweeter) that share a single loudspeaker channel due to hardware crossovers. A total of 17 loudspeakers are used, controlled via 11 independent loudspeaker channels. A rigid spherical microphone array is positioned near the driver's head location during the measurements.}
    \label{fig:loudspeaker arrangement}
\end{figure}
\begin{figure}
    \centering
    \includegraphics[width=0.5\linewidth]{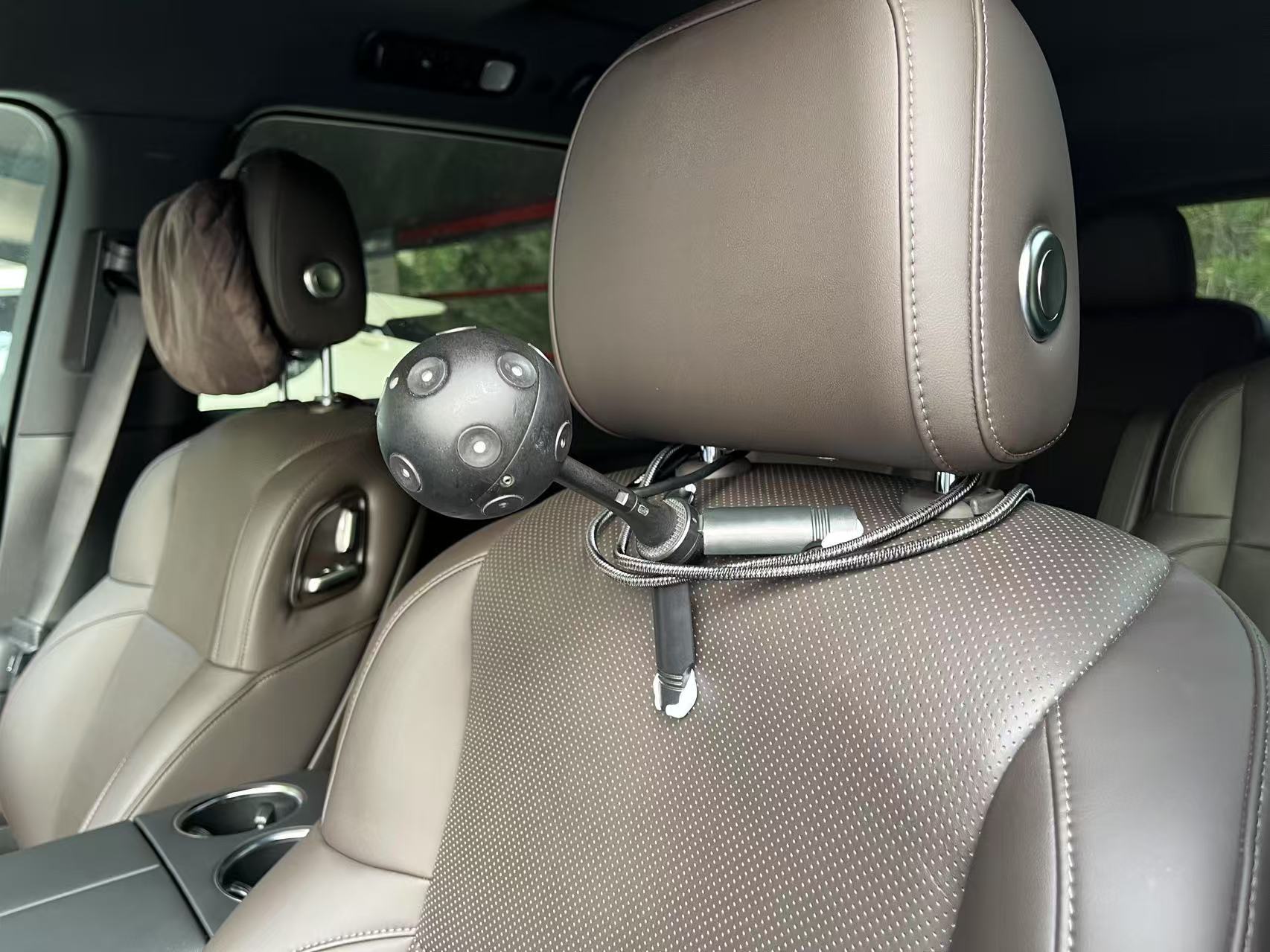}
    \caption{The experimental automotive cabin with an SMA mounted at the position of driver's head (position $O$)}
    \label{fig:SMA_in_car}
\end{figure}

All data measurements and experiments were conducted from the driver’s position in the Li Auto L9 automotive cabin, with a sampling frequency of  $f_s = 48\text{kHz}$. Fig.\ref{fig:loudspeaker arrangement} illustrated the arrangement of the loudspeakers on the horizontal plane within the cabin. A total of 17 physical loudspeakers were installed in the cabin. However, due to analog hardware crossovers, multiple drivers (e.g., tweeter, midrange, woofer) were wired together and cannot be controlled independently. We defined a loudspeaker channel as a digital audio output path that can be independently processed and filtered, following \cite{chen2015spatial}. As a result, only 11 independent loudspeaker channels were available for sound field control. 
The “surround” loudspeakers were positioned at the top, and the “woofer” speakers were located at the bottom of the doors. The remaining loudspeakers were placed slightly below head height, which was typical for most listeners. As depicted in Fig.\ref{fig:SMA_in_car}, a rigid spherical microphone array (SMA) with 3cm radius, consisting of 16 omnidirectional microphones, was positioned at the driver’s head location to measure the acoustic channel within the automotive cabin. The center of the SMA was positioned $58.6\text{cm}$ above the seat surface, located along the vertical plane of the seat. We named after this position as $O$ and positions $4\text{cm}$ and $7.5\text{cm}$ left or right from it as $L/R$ and $LL/RR$. These added positions were introduced to test and enhance the spatial robustness of the controlled sound field. 

\begin{figure}
    \centering
    \includegraphics[width=0.5\linewidth]{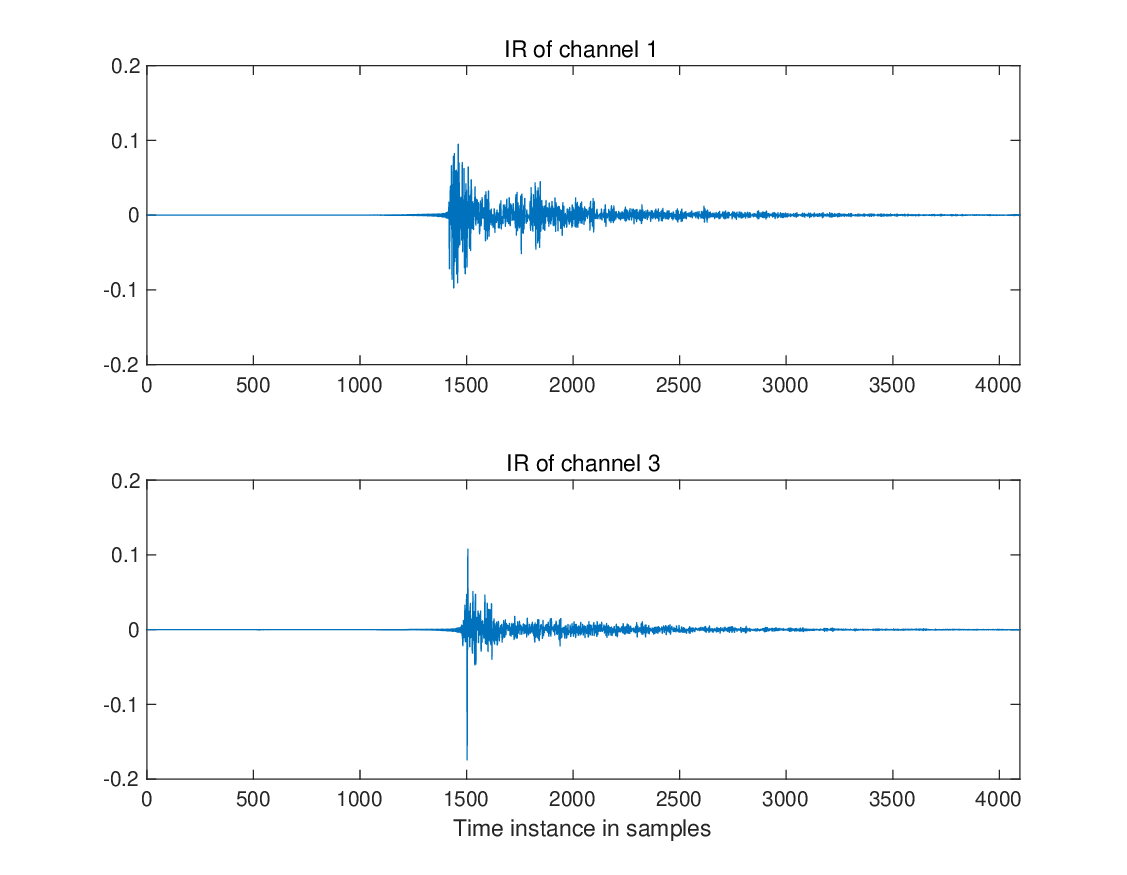}
    \caption{Measured IRs of channel 1 (above), comprising of left front woofer, mid-range, and tweeter together, and channel 3 (below), comprising of center loudspeaker alone, are plotted as representative examples.}
    \label{fig:measured_ir}
\end{figure}
IRs were measured separately for each loudspeaker channel using five repetitions of 2-second exponential sine sweeps ranging from 20Hz to 20kHz\cite{Farina_2007}. To ensure temporal alignment across all channels, a maximum length sequence (MLS) was first played through the center loudspeaker prior to each measurement, serving as a synchronization reference \cite{rife1989transfer}.

We obtained $Q \times L$ IRs of length $L_{c} = 4096$ , where $Q = 16, L = 11$, at each of the 5 positions ($LL, L, O, R, RR$) through measurements. Due to the limited operating frequency range of the loudspeakers and microphones, the frequency range of interest was restricted to 200Hz to 12kHz. Accordingly, the measured impulse responses were subjected to band-pass filtering to retain only the components within this range, and the filtered responses were then used as $c_{lq}(t)$. Fig.\ref{fig:measured_ir} showed $c_{l=1,q=1}(t)$ of channel 1 (left front woofer, mid-range, and tweeter together) and $c_{l=3,q=1}(t)$ of channel 3 (center) at position $O$. In comparison, it was evident that the impulse peak of channel 1 was less pronounced, while that of channel 3 was more prominent. The measured IR for channel 1 showed that the three loudspeakers—each operating in a different frequency band—shared the same channel due to their distinct placements and directivities, resulting in a highly complex impulse response. Additionally, since the energy distribution among these loudspeakers could not be adjusted due to the fixed hardware configuration, we were constrained to designing a single filter for the entire channel to control its overall response. Furthermore, the IRs of both channel 1 and channel 3 revealed significant reverberation caused by sound reflections within the cabin, underscoring the acoustic complexity of the experimental environment. 

After the pre-processing of $c_{lq}(t)$ and the design of appropriate filters $h_{ls}(t)$, exponential sinusoidal sweeps were filtered by $h_{ls}(t)$ and played by the corresponding loudspeaker channel to excite the sound field. To obtain a more realistic overall response, we simultaneously drive all channels during measurement, rather than using the individual measurement approach employed for measuring the IR of a single channel. In order to keep the systemic errors close, the original overall system response was also re-measured in the same way.

Before starting the objective evaluation, we first list the full names, abbreviations, and key characteristics of all the methods to be evaluated. The proposed SPMnet method, representing using Equ.~(\ref{eq:new overall objective function}) as objective function and solving through deep optimization. The CVX method, representing using Equ.~(\ref{eq:overall objective function}) as objective function as solving through convex optimization techniques as in \cite{qian2024automotive}. The FD method, representing directly using frequency deconvolution through least square as in \cite{Kirkeby1998}. The ORI method, representing the original automotive audio system. Furthermore, to verify the effectiveness of introducing the SPM alone, we included method NN, using Equ.~(\ref{eq:overall objective function}) while solving through deep optimization. Additionally, to evaluate whether the inclusion of multiple SMA measurements from different positions improves overall performance, we considered SPMnet-3, which included measurements from 3 positions $LL, O, RR$ as independent training samples.
The target IR $\mathbf{d}_{qs}$ was simulated using SMIRgenerator for all methods \cite{Jarrett2012}.

\subsection{\label{sec3sub2: sqAnalyze} Sound Quality Evaluation}
To provide an overall description of the algorithm’s capability of reproducing sound field of virtual sound images with different DOAs, we calculated the control filters and re-measured the overall IRs for a total of 36 horizontal virtual sound images separately, with DOAs $\varphi_s  = 0^{\circ}, 10^{\circ},\cdots,350^{\circ}$, $\theta = 0^{\circ}$, and distance $r = 1m$. Here, $\varphi_s$ is defined in a horizontal polar coordinate system centered at the listener’s head, with $0^{\circ}$ pointing directly in front of the listener and increasing counterclockwise.
\begin{table}[t]
\centering
\caption{
Comparison of nPRQ and SD metrics for all methods.
nPRQ was divided into \textit{pre}- and \textit{post}-ringing components, evaluated at Position~$O$ (“pos~$O$”) and averaged across five positions (“avg”).
SD was computed across six octave bands starting from 225Hz.
The columns labeled “11.3kHz” and “5.65kHz” indicate the averaged SD over all six octave bands (up to 11.3kHz) and the first five octave bands (up to 5.65kHz), respectively. Lower values indicate better performance, with the optimal result highlighted in boldface.
}
\label{tb:nprq_sd_final}
\begin{tabular}{c c c c c c c c c}
\hline
\hline
\multirow{3}{*}{\textbf{Method}} 
& \multicolumn{4}{c}{\textbf{nPRQ}~$\downarrow$} 
& \multicolumn{4}{c}{\textbf{SD}~$\downarrow$} \\
\cline{2-9}
& \multicolumn{2}{c}{pre} & \multicolumn{2}{c}{post} 
& \multicolumn{2}{c}{11.3kHz} & \multicolumn{2}{c}{5.65kHz} \\
\cline{2-9}
& pos~$O$ & avg & pos~$O$ & avg & pos~$O$ & avg & pos~$O$ & avg \\
\hline
ORI      & 6.08 & 6.25 & 1.43 & 1.34 & 3.21 & 3.06 & 3.15 & 3.07 \\
\hline
FD       & 8.11 & 7.80 & 4.73 & 4.45 & 3.43 & 3.29 & 3.13 & 3.03 \\
\hline
CVX      & 2.86 & 2.46 & \textbf{1.20} & \textbf{0.98} & \textbf{3.00} & 3.05 & 2.66 & \textbf{2.67} \\
\hline
NN       & 2.80 & 3.16 & 1.51 & 1.64 & 3.02 & \textbf{3.04} & \textbf{2.64} & 2.74 \\
\hline
SPMnet   & 2.61 & 2.87 & 2.06 & 2.15 & 3.17 & 3.16 & 2.91 & 2.96 \\
\hline
SPMnet-3 & \textbf{1.79} & \textbf{1.92} & 1.51 & 1.74 & 3.15 & 3.13 & 2.92 & 2.92 \\
\hline
\hline
\end{tabular}
\end{table}

Time and frequency domain metrics were both included to comprehensively analyze phase and amplitude characteristics of the overall responses, providing a basis for objectively judging the sound quality of the reproduced sound field. Normalized perceivable reverberation quantization (nPRQ) \cite{Jungmann_Mazur_Kallinger_Mei_Mertins_2012} was introduced to quantitatively describe the achieved equalization, with separate metrics for pre-ringing artifacts ($\text{nPRQ}_{\text{pre}}$) and post-ringing artifacts ($\text{nPRQ}_{\text{post}}$) in the time domain. nPRQ focuses on the average energy overshoot over all the overshooted time instances. A low nPRQ indicates that the overshoot instances were not severe. The reason for this separation was that the control filter we designed was mixed-phase, which led to the introduction of both types of artifacts. The human auditory system, however, exhibits differing sensitivities to two artifacts \cite{Bücklein_1981}. $\text{nPRQ}_{\text{pre}}$ and $\text{nPRQ}_{\text{post}}$ were calculated as 
\begin{equation}\label{eq:nPRQ}
\begin{aligned}
    \mathrm{nPRQ}  = \left\{ \begin{matrix} 
      &\frac{1}{\left \| \mathbf{g} _{E} \right \|_{0} } \cdot {\textstyle \sum_{n=N_{post}}^{L_g-1}} g_E(n), &\mathrm{for} \ \left \| \mathbf{g} _{E} \right \|_{0} > 0, \ postring\  \\
      &\frac{1}{\left \| \mathbf{g} _{E} \right \|_{0} } \cdot {\textstyle \sum_{n=0}^{N_{pre}}} g_E(n), &\mathrm{for} \ \left \| \mathbf{g} _{E} \right \|_{0} > 0, \ prering\  \\
      &0, &\mathrm{otherwise}.
    \end{matrix}\right.
\end{aligned}
\end{equation}
with $\left \| \mathbf{g} _{E} \right \|_{0}$ denoting the $l_0$ pseudo norm, which counts the number of nonzero elements of vector $\mathbf{g}_E$. The elements $g_E(n)$ of $\mathbf{g}_E$ were calculated using a dual-threshold condition, which suppresses low-energy components and ensures stability, as follows
\begin{equation}\label{eq:g(t)}
\begin{aligned}
\mathrm{g_E(n)}  = \left\{ \begin{matrix} 
  &20\cdot\mathrm{log}_{10}(\left | g(n) \right |\cdot w_u(n) ),  &\mathrm{for} \left | g(n) \right | > \frac{1}{w_u(n)} \mathrm{ and } \left | g(n) \right | > 10^{-2.5}\\
  &0, &\mathrm{otherwise}.
\end{matrix}\right.
\end{aligned}
\end{equation}
$N_{pre}$, $N_{post}$ in Equ.~(\ref{eq:nPRQ}) and $w_u(n)$ in Equ.~(\ref{eq:g(t)}) were all predefined together within the objective function in Equ.~(\ref{eq:new overall objective function}). 

Table \ref{tb:nprq_sd_final} reported the nPRQ metric across all 36 rendered virtual sound images, evaluated at Position~O and averaged across five positions. The trends of the averaged measures closely follow those observed at Position~O, indicating spatial consistency in temporal artifacts. 
ORI exhibited extremely high pre-ringing artifacts and low post-ringing artifacts, likely due to its lack of temporal alignment. When signals from different loudspeakers arrive at different times, a smaller early peak preceding a stronger one is penalized as pre-ringing, while late-arriving components are less likely to be classified as post-ringing artifacts and thus tend to be underestimated. 
All methods except FD effectively reduced temporal artifacts, thereby improving perceived sound quality. 
The weakness of FD was expected, as it constrains only the magnitude of the frequency response and does not enforce any temporal shaping on the system’s impulse response.
NN and CVX yielded similar performance across all nPRQ metrics, suggesting that deep optimization alone may not provide added benefit over conventional convex optimization in this context. 
Interestingly, SPMnet and CVX exhibited complementary strengths: SPMnet showed lower pre-ringing, while CVX outperformed in post-ringing suppression. Most notably, SPMnet-3 outperformed SPMnet in both pre- and post-ringing metrics, which is counterintuitive. Controlling more spatial positions typically leads to degraded performance due to the limited degrees of freedom (loudspeaker channels). This improvement might be attributed to the spatial power map constraint used in SPMnet-3, which implicitly regulates phase coherence across space via beamforming principles. As more spatial positions are introduced, this phase constraint becomes stronger, potentially leading to better temporal behavior.

The well known Spectral Deviation (SD) was introduced as a frequency domain metric for sound quality\cite{Cecchi2017RRE}. SD focuses on the the flatness of the amplitude response of the controlled sound field. A low SD indicates a flat amplitude response, resulting in better sound quality. It is calculated as 
\begin{equation}\label{eq:SD}
\begin{aligned}
    S_{\mathrm{D}}=\sqrt{\frac{1}{f_{\mathrm{h}}-f_{l}+1} \sum_{i=f_{l}}^{f_{\mathrm{h}}}\left(10 \log_{10}\left|E\left(e^{j \frac{2 \pi}{N_f}i}\right)\right|-D\right)^{2}}, 
\end{aligned}
\end{equation}
where 
\begin{equation}\label{D in SD}
\begin{aligned}
    D=\frac{1}{f_{\mathrm{h}}-f_{l}+1} 
    \sum_{i=f_{l}}^{f_{\mathrm{h}}}\left(10 \log _{10}\left|E\left(e^{j \frac{2 \pi}{N_f} i}\right)\right|\right).
\end{aligned}
\end{equation}
$E(e^{j\frac{2\pi}{N_f}i})$ is the frequency response, $N_f$ is the number of frequency indexes, $f_l$ and $f_h$ are the lowest and highest frequency indexes of the interested frequency band. We computed the SD metric for each octave band using rectangular windows centered at their respective center frequencies (e.g., $f_c = 1$kHz). A total of 6 octave bands were included based on the interested frequency band ($225$Hz–$11.3$kHz).

Table~\ref{tb:nprq_sd_final} reported the SD metrics computed by averaging all six bands($5.65$kHz–$11.3$kHz) and the first five octave bands ($225$Hz–$5.65$kHz), respectively, across 36 rendered virtual sound images. Results are reported both at Position~$O$ and as averages over five listening positions. Improvements were observed in the first five bands, while the enhancement was comparatively limited if taking the sixth band into consideration. At Position~$O$, SD of all methods except FD decreased compared to ORI, indicating that sound quality were generally improved. In particular, CVX gained the lowest SD measure, which may yield superior sound quality than other methods. NN and CVX exhibited similar results in SD, indicating that while deep optimization may not necessarily improve performance, it remains effective in solving the optimization problem. SPMnet performed worse than CVX and NN, revealing that the introduction of spatial power map constraint might lead to sound quality impairment. SPMnet-3 was compatible to SPMnet method in SD measure, suggesting that the introduction of multiple SMA don't lead to sound quality impairment. For the FD method, although theoretically it could achieve excellent SD measure, its poor robustness and the fact that we obtained the overall response of rendered sound field through real-world scenario measurements led to significant performance degradation. However, this also demonstrated the robustness of the other methods mentioned. For the results on the averaged SD across all positions, increases were observed in SPMnet, NN and CVX within the 225Hz–5.65kHz frequency band and in NN and CVX within the 225Hz – 11.3kHz, while such increase was not observed in SPMnet-3. This result suggested that methods using a single SMA for control were indeed less robust compared to SPMnet-3 which employed multiple SMAs. It is worth noting that the SD improvement achieved in our study was less pronounced than reported in previous works \cite{Cecchi2009, qian2024automotive}. This may be attributed to the additional constraints imposed in the time and spatial domains, as well as measurement variability caused by microphone placement and environmental factors. Furthermore, hardware optimizations by the vehicle manufacturer—such as loudspeaker arrangement—might have already improved frequency uniformity. Although these factors lay beyond the scope of this study, they warrant further investigation in future work.

Based on the analysis of sound quality, several general conclusions can be drawn. All methods, with the exception of FD, yielded significant improvements in sound quality. Furthermore, the inclusion of multiple SMA positions effectively maintained these enhancements across several locations. However, the impact of incorporating the SPM constraint on sound quality was not entirely consistent between the time and frequency domains. In the time domain, both pre- and post-ringing artifacts were clearly reduced. In contrast, while the constraint still contributed to frequency-domain performance, its ability to maintain response flatness was somewhat reduced compared to other methods.

\subsection{\label{sec3sub3: dirAnalyze} Spatial Localization Evaluation}
\begin{figure}
    \centering
    \includegraphics[width=0.8\linewidth]{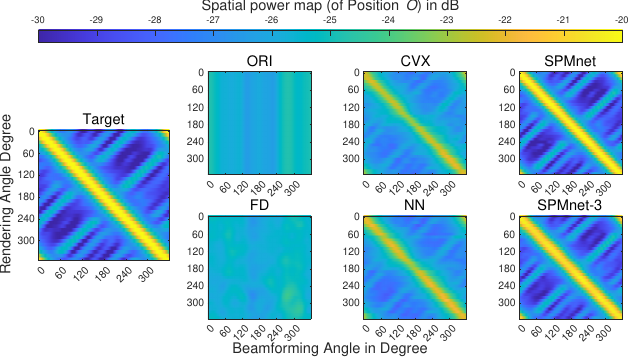}
    \caption{SSPMs of semi-simulated reproduced sound field at Position $O$. Each row corresponds to one virtual source direction, and each column to one beamforming direction. Brighter diagonals indicate stronger spatial localization. }
    \label{fig:semisimu_SSPM}
\end{figure}

To analyze spatial localization performance, stacked spatial power maps (SSPMs), as in Fig.~\ref{fig:semisimu_SSPM}, were computed using delay-and-sum beamforming. A bright main diagonal indicates accurate and focused spatial rendering, where the beamformed direction aligns with the intended virtual source. In contrast, off-diagonal or blurred patterns suggest localization errors or spatial smearing. Horizontal stripes indicate poor energy focus across beamforming angles, while vertical stripes imply that distinct virtual sources are rendered with similar spatial characteristics.

Fig.~\ref{fig:semisimu_SSPM} showed the SSPMs of the semi-simulated reproduced sound field at Position $O$. The semi-simulation was done by convolving the measured IRs with the optimized control filters. The Target showed a bright main diagonal, representing the ideal spatial distribution and serving as the optimization objective. In contrast, the ORI and FD exhibited no clear diagonal structure, indicating poor spatial localization in the original system and the inability of FD to reconstruct directional sound fields. The NN and CVX methods exhibited a visible but relatively faint main diagonal, suggesting that strategies based on relative peak positions and amplitude of impulse responses control can partially enhance localization performance. However, the improvement was limited, and the use of deep optimization alone did not lead to spatial localization improvement. In comparison, both the SPMnet and SPMnet-3 methods display a bright main diagonal, indicating that the proposed algorithms effectively incorporated the spatial power map constraint and significantly improved the spatial localization of the reconstructed sound field.

\begin{figure}
    \centering
    \includegraphics[width=1\linewidth]{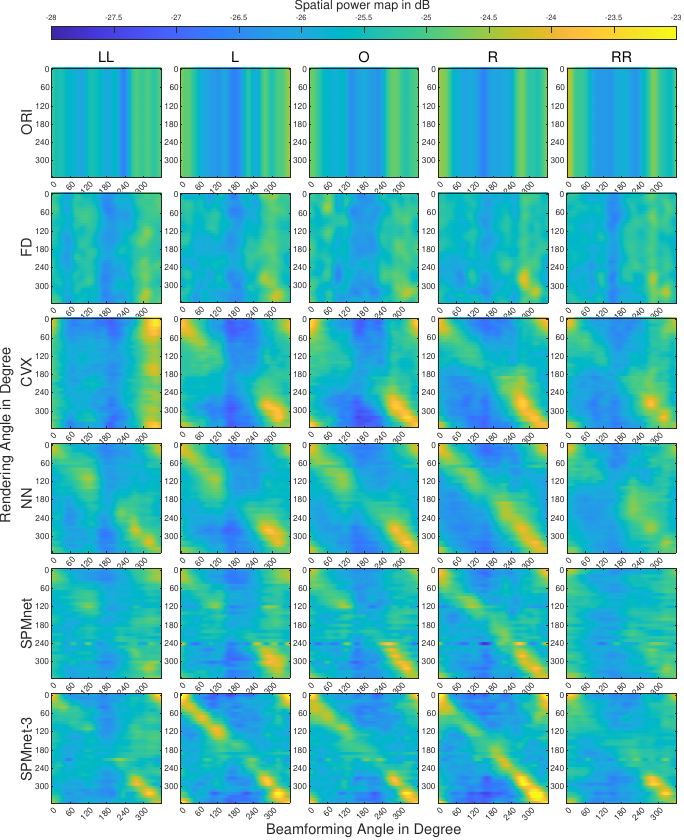}
    \caption{SSPMs of reproduced sound field re-measured in the automotive cabin at all five positions. Each row corresponds to a different method, and each column corresponds to a different listening position.}
    \label{fig:SSPM}
\end{figure}
Fig.\ref{fig:SSPM} showed the SSPMs of the reproduced sound field re-measured in the automotive cabin at all five positions. Compared to the semi-simulated results, the re-measured SSPMs exhibited weaker spatial focus and lower contrast due to the time-varying nature of acoustic channel and measurement variability. Notably, both conditions showed limited dynamic ranges (5–10dB), which were primarily constrained by the spatial resolution of delay-and-sum beamforming and the limited number of microphones—rather than the algorithms themselves. At position $O$, SPMnet-3, SPMnet, NN and CVX methods showed a clear main diagonal in the $240^{\circ}-330^{\circ}$ range, indicating their ability to effectively render virtual sound images at these angles. However, both NN and CVX methods displayed a prominent vertical stripe at the beamforming angle $270^{\circ}$, suggesting that the rendering angles $250^{\circ} - 290^{\circ}$ were difficult to distinguish. From $0^{\circ}$ to $120^{\circ}$(from the front to the left rear side), SPMnet-3 and SPMnet showed an obvious main diagonal, indicating the ability to render these virtual sound images. The SSPMs of NN and CVX were generally comparable. Comparing SPMnet and CVX, we found that the improvements observed in Fig.~\ref{fig:semisimu_SSPM} were less apparent in the re-measured SSPMs—manifested as a thinner, but not significantly brighter, main diagonal. This suggested that while the spatial power map constraint was effective under controlled conditions, its robustness in real-world scenarios may be limited.

To further investigate spatial robustness, we analyzed the results at other positions. Generally speaking, all methods except SPMnet-3 faced ability degradation.
At position $L$, all methods except SPMnet-3 exhibited noticeable performance degradation, with SSPMs appearing more blurred compared to those at position $O$. This was expected, as none of these methods explicitly minimized the spatial power map error at position $L$ during optimization.
In contrast, at position $R$, all methods showed a clear improvement in localization performance. Specifically, previously missing segments along the main diagonal of the SSPMs were filled in, indicating better spatial continuity. This unexpected enhancement may be attributed to the more central and symmetric placement of position $R$ within the cabin, which likely provided more favorable acoustic propagation conditions.
At positions $LL$ and $RR$, the SPMnet method showed a clear decline in performance, with only a faint main diagonal observed—highlighting its limited robustness under spatial variation. In contrast, CVX and NN each exhibited degradation primarily on one side: CVX showed vertical stripes at position $LL$, suggesting poor angular specificity with localization skewed toward $340^\circ$, while NN displayed blurred localization at $RR$. However, both maintained relatively stable performance on the opposite side, indicating some degree of inherent robustness. Notably, SPMnet-3 outperformed the others by maintaining strong diagonal patterns not only at positions $L$ and $R$, comparable to those at $O$, but also showing only moderate performance degradation at both $LL$ and $RR$. This demonstrated the superior spatial robustness of SPMnet-3 across all tested positions and also indicated a larger sweet spot.

In summary, while the SPMnet method showed clear improvements in spatial localization under semi-simulated conditions, its performance degraded noticeably under real-world measurements. This suggests that although the spatial power map constraint effectively enhanced directional rendering, its robustness to acoustic variability remained limited. By contrast, incorporating microphone array measurements at multiple positions substantially improved the spatial consistency of rendering, demonstrating the benefit of multi-position data in enhancing localization robustness.

\section{\label{sec4: subEval} Subjective Experiment}
\subsection{\label{sec4sub1: expPara} Experimental Setup}
Two in-situ listening tests were conducted in the same automotive cabin as in the last section to perceptually evaluate the methods under investigation. To ensure ecological validity and eliminate potential artifacts, playback was delivered directly through the vehicle's built-in loudspeaker system rather than via headphone-based auralization. A total of 13 listeners participated and the average age of them was 24 (from 22 to 31). None of them declared to have hearing impairments, 11 of them declared to have previous experience in similar listening tests. Participants were instructed to sit in the driver's seat and maintain their head position within the designated sweet spot—the same location where the IRs were measured—throughout the entire experiment. 

The experiments employed a double-blind listening test paradigm adapted from the MUSHRA format \cite{ITU-R-BS1534}, with the key modification being the use of an external, headphone-delivered reference to provide a stable, high-fidelity anchor for evaluation. Listeners were instructed to evaluate either the sound quality or the spatial localization of the reproduced sound field in each trial, depending on the experimental condition. Sound quality and spatial localization were evaluated using distinct anchor signals tailored to each perceptual attribute. For sound quality, the anchor was a 3.5kHz low-pass filtered signal reproduced simultaneously by all loudspeakers. For spatial localization, the anchor was the original signal played through the real loudspeaker closest to the opposite side of the target sound image, thereby inducing a maximal localization error. 

A crucial aspect of our methodology is the use of an external reference. The in-car audio system, being the system under evaluation, is by its nature incapable of serving as its own ideal reference. Its performance is precisely what we aim to measure, including all its inherent distortions and spatial limitations. Therefore, to provide listeners with a stable and undistorted anchor for judgment, the reference signals were delivered exclusively via a well-equalized headphone. This allowed subjects to fairly weigh the different types of artifacts introduced by each reproduction method against a consistent, high-fidelity standard. Furthermore, we omitted a hidden reference from our design. The purpose of a hidden reference is typically to verify if listeners can detect subtle impairments. In this study, however, the perceptual difference between the ideal headphone reference and the in-situ loudspeaker reproductions was substantial and easily identifiable. A test to confirm this obvious difference would be superfluous and would not validate the listeners' ability to perform the more nuanced task of comparing the different methods. The reference signals were constructed by convolving the anechoic source material with corresponding Head-Related Impulse Responses (HRIRs) from the PKU-IOA database \cite{Qu2009hrtf}.

According to the result of Sec.\ref{sec3: objEval}, SPMnet-3, SPMnet and CVX were chosen for further comparisons in the subjective evaluation. SPMnet-3 was chosen due to its strong performance in objective evaluation. SPMnet was included for ablation purpose, to evaluate the contribution of the proposed spatial power map constraint while excluding effects from multi-position control. CVX was selected as a baseline method, showing moderate but stable performance. NN was omitted because its performance was similar to that of CVX, offering limited additional insight. FD was excluded due to consistently poor performance in objective evaluations. ORI was not included, as our previous study \cite{qian2024automotive} have already demonstrated that CVX significantly outperforms it.

A total of 5 listening materials were included in the whole experiment. Speech, Symphony, Clapper and String were selected from the audio materials used in \cite{Bauer_Vinton_2005} to test the methods under signals including both transient and sustained signals, as well as broad band and narrow band spectra. A noise burst train was otherwise included for its great performance on the experiment of spatial localization\cite{Zhang_Ge_Liu_Wu_Qu_2020}. For sound quality, 6 rendering angles from $30^{\circ}$ to $330^{\circ}$ with a $60^{\circ}$ step, together with 4 materials (Speech, Symphony, Clapper and String) and 4 methods including sound quality anchor were tested. For spatial localization, 12 rendering angles, from $0^{\circ}$ to $330^{\circ}$ with a $30^{\circ}$ step, together with 3 materials (Symphony, Clapper and noise burst train) and 4 methods including spatial anchor were tested. For clarity in the subsequent analysis, we defined the following abbreviations for the experimental factors: methods (MTH), rendering angles (ANG), and listening materials (MAT). 
\subsection{\label{sec4sub2: sqRes} Sound Quality Evaluation}
\begin{figure}
    \centering
    \includegraphics[width=0.5\linewidth]{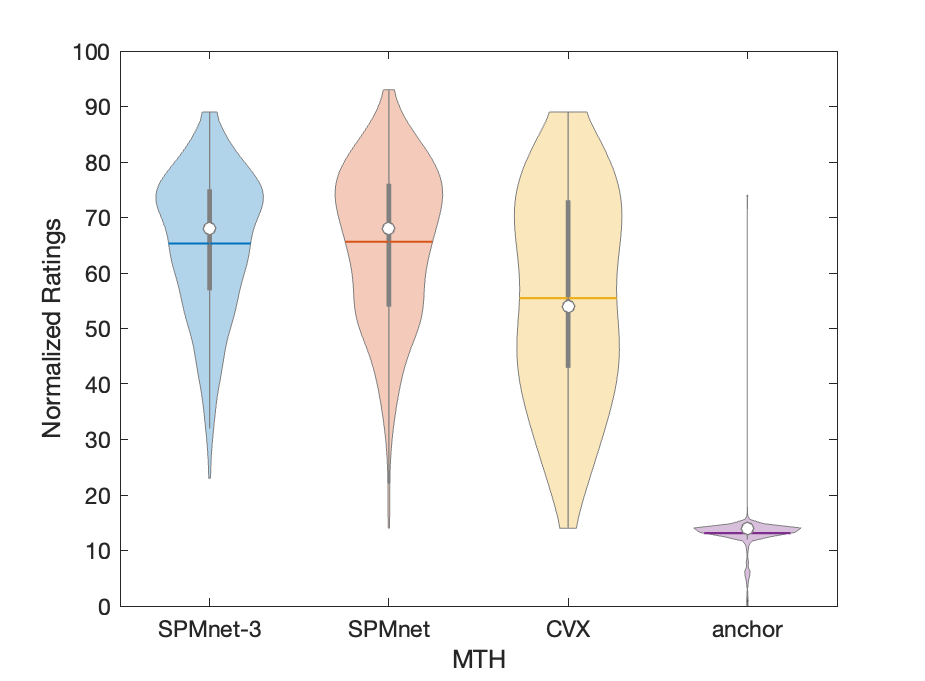}
    \caption{Results from sound quality experiment represented by violin plots that show the probability density of the data, median(circle), interquartile range(box) and mean(horizontal line). }
    \label{fig:sub_timbral}
\end{figure}
Results of the sound quality experiment were shown in Fig.\ref{fig:sub_timbral}. Data for 12 of 13 listeners passed the post-screening based on anchor selection. The displayed data were normalized between the range from 0 to 100 for better visualization but all inferential analysis was performed on the original data. Descriptive analysis showed that data distribution of SPMnet-3 and SPMnet were generally more dense on higher ratings. Mean and median ratings revealed that the order of descending ratings was SPMnet, SPMnet-3, CVX and the Anchor. 

Inferential analysis was performed through repeated measures analysis of variance (RM-ANOVA). The main dependent variable was the reproduction method , its interactions with material and rendering angle were also studied. Since the data did not pass the Mauchly sphericity test ($p<0.001$) and the Greenhouse-Geisser epsilon was higher than 0.75 ($\epsilon = 0.779$), Huynh-Feldt correction was used and a significance level of $\alpha = 0.05$ was used, as recommend in \cite{ITU-R-BS1534}.

(1) Effect of method: A significant effect of MTH was found [$F(2.566,\ 677.425) = 1244.462, p<0.001$]. Post hoc multiple dependent t-tests showed that except between SPMnet-3 and SPMnet [$t(287)=-0.343, p=0.732$], significant differences were found between all pairs of conditions. For the comparing order of ANCHOR, Cvx, SPMnet-3/SPMnet, the results [$ t(287) <= -7.921,\ p < 0.0001 $] indicated that the overall sound quality of MTH were increasing with the comparing, with SPMnet-3/SPMnet outperforming other methods.

(2) Effect of Material: A significant effect was found for the interaction between MAT and MTH $[F(7.698,\ 677.425) = 2.383,\ p = 0.017]$. Dependent sample t-tests between SPMnet-3 and Cvx on all four MAT were used, while the SPMnet was not included for its non-significant difference compared to SPMnet-3.
The results of t-tests were listed in Table \ref{tb:Timbral-ttest}. The results indicated that SPMnet-3 was significantly better than CVX on all materials. The effect of material will be further discussed in Sec.\ref{sec5: discuss}, together with the comparison with objective evaluations.
\begin{table} 
\centering
\caption{Results of post hoc dependent sample t-tests with methods: A:SPMnet-3, B:CVX. Positive t values with $p < 0.05$ indicate that the former method significantly outperforms the latter. Results on all materials appeared to be statistically significant.}
\label{tb:Timbral-ttest}
\begin{tabular}{ccccc}
\hline
\hline
MAT   & Speech & Synphony & Clapper & String  \\ 
\hline
t(71) & 5.727  & 2.713    & 2.391   & 4.975   \\ 
\hline
p     & 0.001  & 0.008   & 0.019   & 0.001   \\ 
\hline
\hline
\end{tabular}
\end{table}

(3) Other interactions: No significant interactions were found between the MTH and ANG [$F(12.830,\ 677.425) = 1.164,\   p = 0.303$] and their three-way interactions with MAT [ $F(38.490,\ 677.425) = 0.786,\ p = 0.821$].

\subsection{\label{sec4sub3: dirRes} Spatial Localization Evaluation}
\begin{figure}
    \centering
    \includegraphics[width=0.5\linewidth]{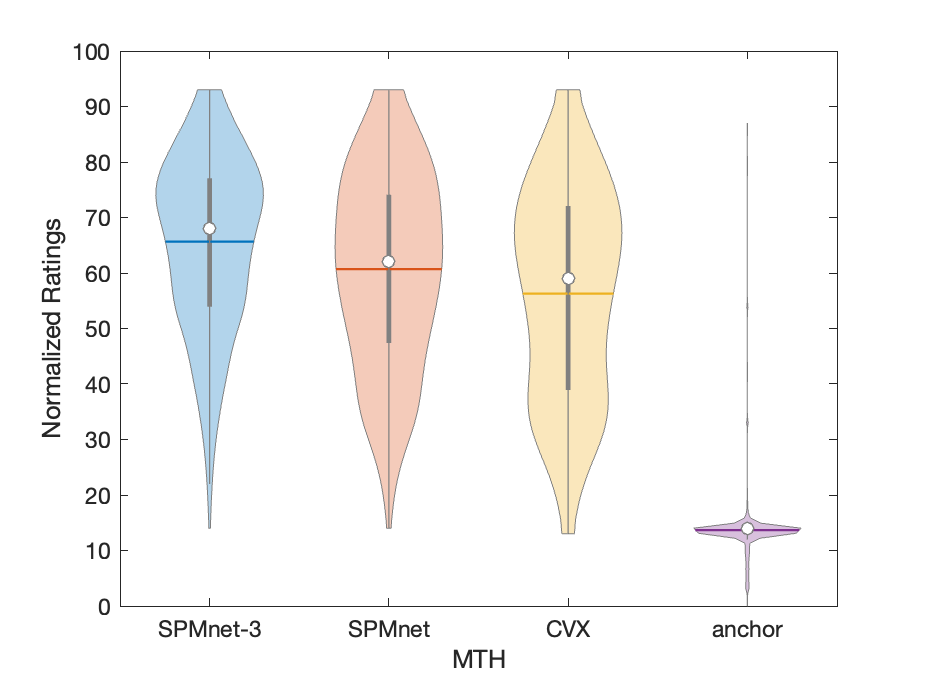}
    \caption{Results from spatial localization experiment represented by violin plots that show the probability density of the data, median(circle), interquartile range(box) and mean(horizontal line). }
    \label{fig:sub_spatial}
\end{figure}

Results of the spatial localization experiment were shown in Fig.\ref{fig:sub_spatial}. Data for 12 of 13 listeners (not the same 12 in the experiment of sound quality evaluation) passed the post-screening. Normalization on data was also done as described in Sec.\ref{sec4sub2: sqRes}. Descriptive analysis showed that data distribution of SPMnet-3 was generally more dense on higher ratings. Mean and median ratings revealed that the order of descending ratings was SPMnet-3, SPMnet, CVX and the Anchor.

Inferential analysis was performed through RM-ANOVA. The main dependent variable was the MTH, but its interactions with ANG and MAT were investigates as well. Since the data did not pass the Mauchly sphericity test ($ p < 0.001 $) and the Greenhouse-Geisser epsilon was higher than 0.75 ($\epsilon = 0.965$), Huynh-Feldt correction was used as recommended\cite{ITU-R-BS1534}. A significance level of $\alpha = 0.05$ was used.

(1) Effect of method: A significant effect of MTH was found on the MUSHRA ratings [$F(3,\ 1188) = 1428.046,\ p < 0.001$]. Post hoc multiple dependent sample t-tests showed that significant differences were found between all pairs of conditions. For the comparing order of ANCHOR, CVX, SPMnet, SPMnet-3, the results were [$t(431) <= -4.26,\ p < 0.001$]. It indicated that the overall spatial localization of the methods were increasing with the comparing order.

\begin{table}[!ht]
\centering
\caption{RM-ANOVA over SPMnet-3, CVX and SPMnet methods. ANG was used as the main effect. Suitable correction were applied according to the Mauchly sphericity test and Greenhouse-Geisser epsilon. Statistically significant results are presented in boldface.}
\label{tb:Spatial-rmANOVA-ANG}
\resizebox{\textwidth}{!}{
\begin{tabular}{ccccccccccccc}
\hline
\hline
ANG     & 000    & 030   & 060    & 090   & 120   & 150   & 180    & 210    & 240    & 270    & 300    & 330     \\ 
\hline
F & \textbf{15.453} & 0.497 & \textbf{12.070} & \textbf{3.492} & 1.625 & 0.171 & \textbf{15.876} & \textbf{11.887} & \textbf{13.139} & \textbf{12.678} & \textbf{32.207} & \textbf{15.465}  \\ 
p       & \textbf{$<$0.001}  & 0.611 & \textbf{$<$0.001}  & \textbf{0.036} & 0.205 & 0.835 & \textbf{$<$0.001}  & \textbf{$<$0.001}  & \textbf{$<$0.001}  & \textbf{$<$0.001}  & \textbf{$<$0.001}  & \textbf{$<$0.001}  \\
\hline
\hline
\end{tabular}
}

\end{table}

(2) Effect of Angle: A significant effect was found for the interaction between ANG and MTH $[F(33,\ 1188) = 6.292,\ p < 0.001]$. Given that a large number of dependent sample t-tests will lead to high Type I error, we turned to use RM-ANOVA for each ANG only among SPMnet-3, Cvx and SPMnet, which is also the comparing order of the post hoc t-tests, for simplicity. The results were listed in Table.\ref{tb:Spatial-rmANOVA-ANG}. Significant difference was found in all ANGs other than $030, 120$ and $150$ $[ F(3,\ 99) >= 11.887,\ p < 0.001]$. Post hoc multiple dependent sample t-tests were run on these ANGs and the results were listed in Table.\ref{tb:Spatial-ttests}. The comparisons between CVX and SPMnet-3 or SPMnet both indicated that on ANGs with significant difference, CVX was the one with worse performance. The comparisons between SPMnet-3 and SPMnet revealed that on ANGs ($060, 090, 240, 300$) with significant difference, SPMnet-3 performed better. The effect of angle was further discussed in Sec.\ref{sec5: discuss}, together with the comparison with objective evaluations.

\begin{table}
\centering
\caption{Results of post hoc dependent sample t-tests with methods: A:SPMnet-3, B:SPMnet and C:CVX. Positive t values with $p < 0.05$ indicate that the former method significantly outperforms the latter. Statistically significant results are presented in boldface. For example, comparison on ANG 000 [t(35); A--C = 4.555 > 0, p < 0.001] demonstrated that A.SPMnet-3 performs significantly better than C.CVX on ANG 000.}
\label{tb:Spatial-ttests}
\resizebox{\textwidth}{!}{
\begin{tabular}{ccccccccccccc}
\hline
\hline
ANG & 000 & 030 & 060 & 090 & 120 & 150 & 180 & 210 & 240 & 270 & 300 & 330 \\
\hline
$t(35;\ \text{A--B})$ 
& -1.209 & 0.757 & \textbf{5.785} & \textbf{2.727} & -1.229 & -0.622 & 1.782 & -1.982 & \textbf{4.864} & -0.337 & \textbf{7.043} & 1.631 \\
$p(\text{A--B})$ 
& 0.235 & 0.454 & \textbf{$<$0.001} & \textbf{0.010} & 0.227 & 0.538 & 0.084 & 0.055 & \textbf{$<$0.001} & 0.738 & \textbf{$<$0.001} & 0.112 \\
\hline
$t(35;\ \text{A--C})$ 
& \textbf{4.555} & 0.812 & \textbf{2.474} & 1.146 & -1.763 & -0.293 & \textbf{5.237} & \textbf{3.090} & \textbf{3.509} & \textbf{4.599} & \textbf{5.450} & \textbf{6.033} \\
$p(\text{A--C})$ 
& \textbf{$<$0.001} & 0.422 & \textbf{0.018} & 0.259 & 0.087 & 0.771 & \textbf{$<$0.001} & \textbf{0.004} & \textbf{0.001} & \textbf{$<$0.001} & \textbf{$<$0.001} & \textbf{$<$0.001} \\
\hline
$t(35;\ \text{B--C})$ 
& \textbf{4.234} & 0.318 & -1.780 & -1.335 & -0.254 & 0.299 & \textbf{3.254} & \textbf{4.578} & -0.869 & \textbf{3.995} & -1.609 & \textbf{3.475} \\
$p(\text{B--C})$ 
& \textbf{$<$0.001} & 0.752 & 0.084 & 0.190 & 0.801 & 0.767 & \textbf{0.003} & \textbf{$<$0.001} & 0.391 & \textbf{$<$0.001} & 0.117 & \textbf{0.001} \\
\hline
\hline
\end{tabular}
}
\end{table}

(3) Other interactions: No significant interactions were found between the MTH and MAT [ $F(6,\ 1188) = 1.645,\ p = 0.131$] and their three-way interactions with ANG [ $F(66,\ 1188) = 1.282,\ p = 0.068$].

\section{\label{sec5: discuss} Discussion}
\subsection{\label{sec5sub1: sqDis} Sound Quality Evaluation}
For the discussion of sound quality, we specifically focused on the differences between time-domain and frequency-domain metrics, as presented in Sec.\ref{sec3sub2: sqAnalyze}, and sought to interpret them in conjunction with subjective evaluations. Furthermore, we examined the material-dependent variations discussed in Sec.\ref{sec4sub2: sqRes} and analyzed the influence of pre-ringing artifacts introduced by the system.

As mentioned in Sec.\ref{sec4sub2: sqRes}, significant differences were observed between SPMnet-3/SPMnet and CVX. Overall, SPMnet-3/SPMnet consistently outperformed CVX under all conditions. However, these results did not directly align with time-domain or frequency-domain evaluation metrics but instead exhibited a more complex relationship. As shown in Table \ref{tb:nprq_sd_final}, CVX suffers from a relatively higher $\text{nPRQ}_{\text{pre}}$ compared to SPMnet-3, indicating a greater presence of pre-ringing artifacts. Although both $\text{nPRQ}_{\text{post}}$ and SD suggest that CVX performs better, the excessive pre-ringing artifacts introduced led to degraded sound quality in subjective evaluations. 

Moreover, the influence of materials can also be interpreted through the $\text{nPRQ}_{\text{pre}}$ metric. Compared to the other three materials, Symphonic exhibits weaker temporal transients, making it less sensitive to time-domain distortions caused by pre-ringing and post-ringing artifacts. In contrast, Speech and String materials contain stronger temporal transient structures, which made such distortions more perceptible  to most participants, resulting in smaller perceptual differences across methods. Interestingly, although Clapper rendered by CVX was identified as having the worst sound quality in a preliminary study, some participants in the subjective experiment reported that even noted that CVX achieved better sound quality despite the presence of audible pre-ringing artifacts. The results in Table \ref{tb:Timbral-ttest} further supported this observation, showing a relatively low effect size ($t(71)=2.391$) for Clapper compared to the other three materials. This phenomenon may be attributed to the interaction between participants’ hearing sensitivity and the frequency components of the material \cite{liski2021}, which warrants further investigation.

These findings demonstrated a key benefit of our approach: the introduction of the DSB-based SPM constraint not only preserves but often enhances the perceived sound quality compared to the CVX method. This improvement is primarily achieved by mitigating the audible pre-ringing artifacts that degrade the CVX solution, despite its superior performance on some objective metrics. This also suggests that for the practical application of non-minimum phase filters in sound field reproduction, more attention should be paid to pre-ringing artifacts, as emphasized in \cite{Brannmark_Ahlen_2009, maeda2021sound}. Furthermore, the control of pre-ringing artifacts may need to be optimized differently for various types of materials to achieve the best perceptual performance.

\subsection{\label{sec5sub2: dirDis} Spatial Localization Evaluation}
The discussion on spatial localization focused on the differences in rendered angles of various methods and whether these differences remained consistent across subjective and objective evaluations. This analysis also helped assess the effectiveness of DSB-based SPM constraint in enhancing the spatial localization of the reproduced sound field.

As illustrated in Sec.\ref{sec4sub3: dirRes}, significant differences were observed among SPMnet-3, SPMnet, and CVX, with performance ranking in descending order. In general, this result aligned well with the objective evaluation results shown in Fig.\ref{fig:SSPM}, where a brighter diagonal was observed for SPMnet-3, followed by SPMnet and then CVX at positions L, O, and R.

As an ablation study, we first investigated the differences between SPMnet and CVX. According to Table \ref{tb:Spatial-ttests}, SPMnet achieved better spatial localization at rendering angles of $0^\circ, 180^\circ, 210^\circ, 270^\circ$, and $330^\circ$, which also aligned well with the SSPM shown in Fig.\ref{fig:SSPM}, particularly at Position R. Notably, CVX lacked a diagonal pattern at positions L, O, and R for angles $180^\circ$ and $210^\circ$, indicating its drawback to render sources behind the listener. Meanwhile, SPMnet exhibited narrower lateral ripples at the remaining angles, demonstrating stronger spatial localization for sound sources arriving from these directions. This result strongly supported the effectiveness of introducing DSB-based SPM constraint.

Next, we compared SPMnet-3 and SPMnet. As indicated by the subjective experiment, SPMnet-3 demonstrated better spatial localization than SPMnet at angles $60^\circ, 90^\circ, 240^\circ$, and $300^\circ$. In the objective experiment, SPMnet-3 exhibited a clearer main diagonal in the upper-left ($0^{\circ}-120^{\circ}$) and lower-right ($250^{\circ}-360^{\circ}$) regions across all $L$, $O$, and $R$ positions, which were consistent with the subjective evaluations. Further, only SPMnet-3 exhibited a clear main diagonal in the lower-right at positions $LL$ and $RR$. However, at $30^\circ$, despite SPMnet-3 displaying narrower and brighter lateral ripples, the subjective experiment revealed no significant difference between SPMnet-3, SPMnet, and even CVX. We speculated that this might have been due to the experiment being conducted in the driver’s seat, where the left side was relatively narrow and close to the car window, whereas the right side was more spacious and lacked nearby reflective surfaces. This speculation aligned well with the report from the participants that "Sound sources at the left-front side are generally harder to localize than others".

These results illustrated that the introduction of DSB-based SPM constraint and multiple SMA measurements improved the spatial localization of the reproduced sound field. Furthermore, the SSPMs served as an effective evaluation for spatial localization, as it closely aligned with the results of subjective evaluations. However, both the significant rendering variations across different sound source directions and the interaction between the loudspeaker arrangement and the environment warrant a more detailed analysis.

\section{\label{sec6: conclusion} Conclusion}
In this paper, we proposed SPMnet, a novel method for automotive sound field reproduction. To address the challenging in-cabin acoustic environment, SPMnet introduces a unified sound field reproduction framework that integrates acoustic channel equalization with virtual sound source rendering, thus improving both sound quality and spatial localization. A key contribution is a spatial domain constraint, the DSB-based SPM constraint, which was designed to explicitly enhance spatial localization. To tackle the non-convexity arising from this constraint, we employed a deep optimization approach. This technique leverages deep learning to solve the optimization problem, thereby relaxing the convexity requirement that has limited traditional methods.

We conducted experiments in a real automotive cabin, where both objective metrics and subjective listening tests consistently validated the effectiveness of the proposed method. The results demonstrated not only accurate spatial localization of the reproduced sound field but also significant improvements in overall sound quality. Furthermore, by analyzing the consistency and discrepancies between objective and subjective results, we found evidence supporting the human auditory system's limited sensitivity to pre-ringing artifacts. This finding suggests that pre-ringing control strategies can be adaptively tuned based on the temporal characteristics of the source material.

Future work will pursue two primary directions. The first is to extend the sound field reproduction to a larger listening area, enabling coverage for multiple seats simultaneously. The second is to develop techniques for the independent rendering of distinct sound sources to multiple listening positions, aiming to deliver unique spatial cues to each listener without interference.

\begin{acknowledgments}
This work is supported by the National Key Research and Development
Program of China (No.2024YFB2808902), and the High performance Computing
Platform of Peking University.
\end{acknowledgments}

\noindent\section*{AUTHOR DECLARATIONS}
The data that support the findings of this study are available from the corresponding author upon reasonable request. The study and methods followed the tenets of the Declaration of Helsinki. Both the consent procedure and the experimental procedures of this study involving human participants were approved by the Committee for Protecting Human and Animal Subjects at Peking University. Data acquisition followed the general data protection regulation. The authors confirm that there are no conflicts of interest in the presented work.

\newpage
\bibliography{ref}
\end{document}